# Enhanced transmission capacity for laser communication at the single-photon level using the multi-channel frequency coding scheme

Jianyong Hu, Bo Yu, Mingyong Jing, Liantuan Xiao* and Suotang Jia
*State Key Laboratory of Quantum Optics and Quantum Optics Devices, Institute of Laser Spectroscopy, Shanxi University, Taiyuan 030006, China*
*Collaborative Innovation Center of Extreme Optics, Shanxi University, Taiyuan 030006, China;*
*Corresponding author E-mail address: xlt@sxu.edu.cn

The statistical properties of a radiation sources are commonly characterized by second-order-correlation or Mandel parameter. Our research found that the single photons modulation spectrum provides us another optional way which is more sensitive to the high frequency information contained in the photon sequence. In this paper, we present direct laser communication by using a multi-channel frequency coding scheme based on the single photons modulation spectrum in which the multi-frequency modulation makes the transmission capacity efficiently enhanced. The modulation frequencies could be operated in a wide band without frequency aliasing due to the inherent randomness of photons arrival time of weak coherent light. The error rate less than $10^{-5}$ has been achieved experimentally when the mean signal photon count is 80 kcps. The modulated coherent light field shows nonlinear effects of single photons modulation spectrum. The studies of statistical properties of the single photons modulation spectrum, including the dependence of mean noise photon count, integration time, channel spacing and the number of frequency component, helped us to optimize the error rate and transmission capacity.
**Keywords:** Communications, single photons, transmission capacity, multi-channel frequency coding, error rate.
**PACS:** 03.67.-a, 05.30.-d, 42.79.Sz, 42.81.Uv

## I. INTRODUCTION

Laser communication has been extensively used in the daily life. Commonly, intense laser could be received at the receiving terminal, in which the error rate are remaining at a low level. However, in some extreme cases, such as the satellite-to-ground communication and interplanetary Internet, due to the harsh channel environment or long transmission distance, the received light is weakened to as low as single-photon level [1, 2]. Additionally, the scattering, rapids, thermal blooming effect of the atmosphere would cause the high background noise [3-5].

In another case of the quantum communication, to use the quantum properties of single-photon, the laser intensity has to be man-made attenuated to the single-photon level to guarantee the security of communication [6-11]. Thereinto, imperfection of devices and attack operations of the eavesdropper would lead to the noise. The channel loss and noise would lead to a high error rate.

Although the second-order-correlation and Mandel parameter usually be used to characterize the statistical properties of the optical field, however, they are not suitable for communication due to their less degrees of freedom and small signal-to-noise ratio. In reality, optical homodyne phase-shift-keying modulation, pulse position modulation and preamplified differential phase shift keying are commonly used coding scheme in the average-power-limited photon-starved links [12-14]. In these schemes, photon counting is used to detect weak light signal [15-17]. In an ideal photon counting measurement, the signal-to-noise ratio (SNR) is strongly dependent on the mean signal photon count, the highest SNR of coherent light is $N^{1/2}$ (where $N$ is the mean signal photon count within an integration time) which is limited by the quantum shot noise. Additionally, the photon counting is sensitive to the noise photon counts and dark counts. Then photon counting modulation technology was developed to improve the SNR of the single photons detection [18-22]. However, confined by the slow processing speed of photon counting, the modulation frequency is usually less than 10 Hz which directly lead to low transmission capacity.

Recently, the single photons modulation spectrum (SPMS) was developed to research the statistical properties of a radiation sources in frequency domain [23, 24]. In this paper, a multi-channel frequency coding (MCFC) scheme is proposed in which the SPMS is extended to the multi-frequency components which make it efficient for information transmission. The statistical properties of the SPMS, such as nonlinear effects and time dependence, were researched to optimize the error rate and transmission capacity. Finally, transmission capacity is presented considering the error rate. The SPMS could not only be used in the communication, but also in the studies of molecular dynamics, properties of single photon source and thermal optical field, and so on.

## II. PRINCIPLE OF MCFC SCHEME

Photon count of coherent light $|\alpha\rangle$ at the single-photon level obeys Poisson distribution $p_n$ which shows as white noise in the frequency domain [24, 25]. When the emission probability of single photons of coherent light is modulated by an intensity modulator with a cosine function, the distribution change to the superposition of the cosine distribution and the Poisson distribution in the time domain [26-28],

$$P_{out} = \sum_{n=0}^{\infty} p_n \cdot \frac{|\alpha|^2}{2} \cdot [1+\cos(\theta)]. \quad (1)$$

Here, $\theta$ is the relative phase which is proportional to the voltage applied on the Mach-Zehnder interferometer type optical intensity modulator. The distribution changes of modulated optical field are reflected in second-order-correlation function or Mandel Q parameter [29-31], but the modulation frequency can't be read out directly from them. (A detailed description and measurement result of second-order-correlation and Mandel Q parameter can be found in Sec. S1 of the Supplemental Material [32].) The spectrum of the modulated optical field is changed to the superposition of the white noise and a characteristic spectrum line. The modulation frequency can be extracted from the spectrum directly by locating the position of the characteristic spectrum line [33].

Poisson distribution is an intrinsic property of weak coherent light, in principle, photons are randomly detected by the single-photon detector. According to the non-uniform sampling theory, the modulation frequency can be operated in a wide frequency bandwidth without frequency aliasing [34].

Assume that the receiver detected $N$ single photons in the integration time $T_I$ (i.e. the time length of the photon sequence). The arrival time of each photon is $\tau_1, \tau_2, \ldots, \tau_i, \ldots, \tau_N$ ($0 \leq \tau_i < T_I$). For coherent light without modulation, the probability density distribution of photons is $P_{(t)}=1/T_I$, ($0<t<T_I$). The expectation of the Fourier transform of the photon sequence could be expressed as

$$X_{(\omega)} = \sum_{i=1}^{N} \int_0^{T_I} A_{\tau_i} e^{-j\omega\tau_i} \cdot \frac{1}{T_I} \cdot d\tau_i. \quad (2)$$

$A_{\tau_i}=1$ is the output pulse amplitude of single-photon detector. When the photon sequence is modulated with cosine wave function through an intensity modulator, the arrival time of the photons are $\tau_1, \tau_2, \ldots, \tau_l, \ldots, \tau_N$, $N$ obeys Poisson distribution when $T_I \gg 1/f_M$, here $f_M$ is the modulation frequency. The probability density distribution is changed to $P_{(t)}=A(\sin(2\pi f_M t+\varphi)+1)$, ($0<t<T_I$), here $A$ is the normalization coefficient, $\varphi$ is the initial phase of the modulation signal. The expectation of the Fourier transform of the modulated photon sequence is

$$X_{(\omega)} = \int_0^{T_I} A \cdot e^{-j\omega t} \cdot (\sin(2\pi f_M t)+1) \cdot dt$$

$$= -\frac{j(1-e^{j\omega T_I})}{\omega}$$

$$+ \frac{e^{-j\omega T_I} \cdot (-2\pi f_M e^{j\omega T_I} + 2\pi f_M \cdot \cos(2\pi f_M T_I) + j\omega \cdot \sin(2\pi f_M T_I))}{\omega^2 - 4f_M^2 \pi^2}. \quad (3)$$

Here we set $\varphi=0$. The right-hand side of the Eq. (3), the first item is a direct current item, which shows as low-frequency noise. The second item is an alternating current item, which contains the information about the modulation frequency. The Eq. (2) and Eq. (3) indicate that once enough modulated photons are detected, one can get the information of the modulation frequency. By encoding the information on the modulation frequency at the transmitting terminal and decoding the modulation frequency at the receiving terminal, the information can be transmitted.

## III. EXPERIMENTAL SETUP

To demonstrate the MCFC scheme, a typical color image was transmitted. The experimental setup is described in Fig. 1. Each pixel point of the color image could be described by using three parameters: R, G and B (Red, Green and Blue), which denote gray values of three primary monochromatic colors. In the experiment, at the

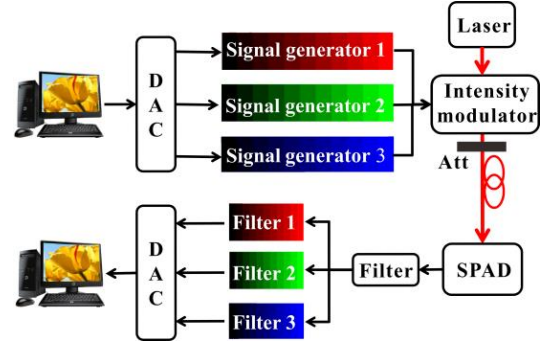

FIG. 1. (color). Experimental setup of the MCFC scheme. DAC: Data acquisition card; Att: Optical attenuator. SPAD: Single-photon avalanche photodiode. The filter was set at 10~100 kHz; the filter1, filter2 and filter3 were set at 60~80 kHz, 40~60 kHz and 20~40 kHz, respectively.

TABLE. I (color). The correspondence between the modulation frequencies and the gray levels of the images.

| Red level | | | | | | | | | | |
|---|---|---|---|---|---|---|---|---|---|---|
| $f_M$ (kHz) | 75 | 74 | 73 | 72 | 71 | 70 | 69 | 68 | 67 | 66 | 65 |
| Green level | | | | | | | | | | |
| $f_M$ (kHz) | 55 | 54 | 53 | 52 | 51 | 50 | 49 | 48 | 47 | 46 | 45 |
| Blue level | | | | | | | | | | |
| $f_M$ (kHz) | 35 | 34 | 33 | 32 | 31 | 30 | 29 | 28 | 27 | 26 | 25 |

transmitting terminal, the original color image was first decomposed into three monochromatic images: Red, Green and Blue respectively. To transmit the image we first need to build up the relationship between the modulation frequencies and the gray levels. The corresponding concrete relationship is listed in Table I. The modulation frequency band is divided to three sub-bands: 60~80 kHz, 40~60 kHz, and 20~40 kHz corresponding to the gray values of red, green and blue respectively. Gray values of each primary color are divided into eleven (0~10) levels, and the channel spacing between the adjacent frequency channels was set at 1.0 kHz. Then three modulation signals with different frequencies were applied on the optical intensity modulator synchronous. Each frequency represents the gray level of one primary color. An optical attenuator was used to simulate the channel loss. At the receiving terminal, only signal at the single-photon level is received. Then a single-photon avalanche photodiode was used for single photons detection. In order to discern the modulation frequencies, electronic filters were used to isolate three sub-bands.

## IV. RESULT AND DISCUSSION

### A. Experimental Results

Figure 2 is the spectrum of the modulated photon sequence when the three modulation signal applied on the modulator simultaneously. From left to right, characteristic spectrum lines locate at 25 kHz, 50 kHz and 71 kHz, respectively. According to the corresponding relationship of the modulation frequencies and the gray levels showed in Table I, these characteristic spectrum lines represent blue 10-th level, green 5-th and red 4-th level, respectively. After the information of all pixels of the image was received, we are finally able to recover the image. The noise floor is white noise which mainly caused by shot noise. First row of Fig. 3 is the original image. The second and third rows show the transmitted images when the mean signal photon count are 80 kcps (kilo counts per second) and 10 kcps, respectively. The integration time $T_I$ is 1 ms, system repetition frequency is 10 MHz, that is, the mean photon per pulse is $8.0 \times 10^{-3}$ and $1.0 \times 10^{-3}$ respectively. There are no errors in the image (b). However, in the image (c), there show a lot of errors in the image when the mean signal photon count is too less to recover the information correctly.

Files with any format can be transmitted using the MCFC scheme by building up the corresponding relationship between the file and the modulation frequency. Such as an article (or a binary file), corresponding relationship between the letters (or the binary sequences) and the modulation frequencies should be established before the transmission. A detailed description is given in the Sec. S2 of the Supplemental Material [32].

### B. Discussion

For a MCFC communication system with a fixed channel spacing $f_s$, the transmission capacity is determined by the total available modulation bandwidth which is mainly limited by the time jitter of SPAD. Assume that there are $M$ frequency channels. Evidently, for a given system there is always a finite optimal value $M_{opt}$ of the number of frequency channels. If $M<M_{opt}$, it is a waste of channel resources. In contrast, if $M>M_{opt}$, the disturbance of adjacent channels would be happened. The optimal number of frequency channels could be calculated by

$$M_{opt} = \frac{B}{f_s} + 1, \quad (4)$$

here $B$ is the total available modulation bandwidth. The combinations of $k$ frequency components make the number of effective frequency channels increase greatly,

$$M_{max} = \frac{M_{opt}!}{k!(M_{opt}-k)!}. \quad (5)$$

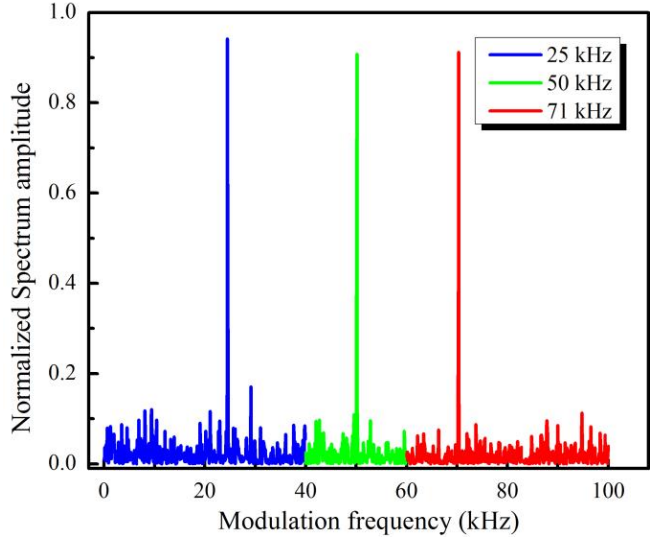

FIG. 2. (color). The spectrum of the modulated photon sequence. The mean signal photon count is 80 kcps, the system repetition frequency is 10 MHz and the integration time is 1 ms.

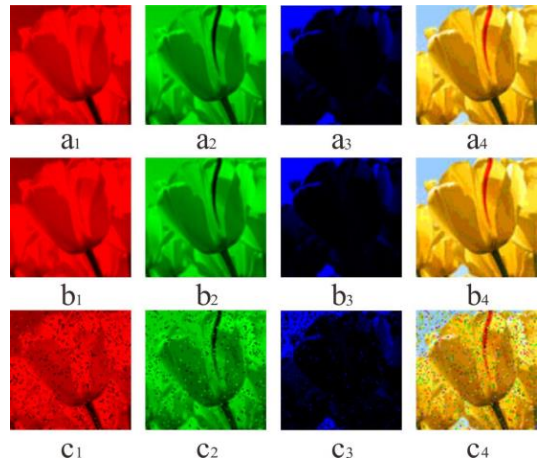

FIG. 3. (color). Image transmission with MCFC scheme. $a_1$, $a_2$ and $a_3$ are the three monochromatic images decomposed from original image $a_4$. $b_1$, $b_2$, and $b_3$ are received monochromatic images when the mean signal photon count is 80 kcps. $b_4$ is the recovered color image by composing the $b_1$, $b_2$, and $b_3$. $c_1$, $c_2$, $c_3$, and $c_4$ are the received images when the mean signal photon count is 10 kcps.

The transmission capacity could be expressed as

$$I = \frac{1}{T_I} \text{Log}_2 M_{max}. \quad (6)$$

The transmission capacity is also influenced by the error rate which mainly depends on the channel loss, noise, channel spacing, frequency component and integration time. Quantitative analysis of error rate is made through numerical simulation. The results are showed in Fig. 4 and Fig. 5.

Quantum shot noise is an intrinsic property of weak coherent light, which would result in the fluctuation of the characteristic spectrum lines. The distribution of the amplitude of the characteristic spectrum line $A_S$ can be characterized through mean value $E[A_S]$ and variance $\sigma_S$. Background light and dark counts of single-photon detector would lead to the increase of noise floor, however, almost no contribution to the characteristic spectrum line. The distribution of the amplitude of the noise floor $A_B$ can be characterized through mean value $E[A_B]$ and variance $\sigma_B$. The amplitude and variance of the characteristic spectrum line and noise floor showed nonlinear effect with the increase of mean signal photon count (as show in the Supplemental Material [32]), which lead to the nonlinear relationship between the error rate and the mean signal photon count. When the amplitude distributions of the characteristic spectrum line and the noise floor are not overlapped, the information can be decoded correctly. As the red line in Fig. 4 shows, the gray dash line D is the boundary of the noise floor. Here the signal photon counts and the noise photon counts both obey Poisson distribution, the mean signal photon count and mean noise photon count is $N_{signal}$=80 kcps and $N_{noise}$=0 kcps, respectively. The noise photon counts will increase the noise floor which may lead to the overlap of amplitude distributions of the characteristic spectrum line and the noise floor. Here we consider a situation that the mean noise photon count is equal to the mean signal photon count, as the black dot line described in Fig.4. Where both of the mean signal photon count and the mean noise photon count is 80 kcps. The black solid line is the sum of the two black dot lines, the boundary (the gray dash line $E$) of the noise floor can be easily found out from it. With the increase of the photon counts, the mean amplitude of characteristic spectrum line $E[A_S]$ increase faster than that of the noise floor. When the mean signal and mean noise photon count are 160 kcps, the noise floor and characteristic spectrum line distribution are split totally, as the blue line showed.

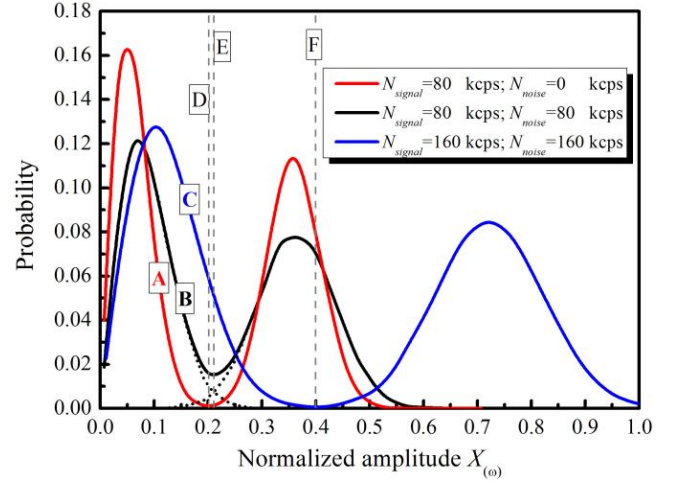

FIG. 4. (Theoretical calculation). The spectrum amplitude distribution of the characteristic spectrum line and the noise floor. The mean signal photon count of red line, black line and blue line are 80 kcps, 80 kcps and 160 kcps, respectively, while the mean noise photon counts are 0, 80 kcps and 160 kcps, and the boundary of noise floor are marked with gray dash line $D$, $E$ and $F$, respectively. The modulation frequency is 200 kHz.

The error rate $e$ is determined together by the amplitude distribution of the characteristic spectrum line and the noise floor. The amplitude of the characteristic spectrum line and the noise floor obeys Gaussian distribution. When the noise amplitude is bigger than the amplitude of the characteristic spectrum line, there will be the error happened. Therefore, the probability of decoding the modulation frequency incorrectly is

$$p = \int_0^\infty \frac{1}{\sqrt{2\pi}\sigma_S} e^{-\frac{(A_S - E[A_S])^2}{2\sigma_S^2}} \left( \int_{A_S}^\infty \frac{1}{\sqrt{2\pi}\sigma_B} e^{-\frac{(A_B - E[A_B])^2}{2\sigma_B^2}} dA_B \right) dA_S. \quad (7)$$

When there are $M$ frequency channels. The error rate is $e=1-(1-p)^M$, the error rate $e$ increases with the number of frequency channels. In our experiment, when the mean signal photon count is 80 kcps, the error rate $e=1\times10^{-5}$. However, the error rate is up to 0.10 when the mean signal photon count is 10 kcps. Parameters which influence the error rate is researched in Fig. 5.

Figure 5(a) researches the error rate of the MCFC scheme in a noisy environment quantitatively. Lines with different color represent different mean signal photon count. It is found that the error rate is maintaining a low level when the mean noise photon count is comparable with the mean signal photon count. As the blue line shows, the mean signal photon count is 160 kcps, while the error rate is still less than $10^{-6}$ when the mean noise photon count is 160 kcps. When the mean noise photon count is zero, the error rate is only limited by the quantum shot noise.

Besides the noise count dependence, the error rate is also correlated with the integration time $T_I$. As show in Fig. 5(b),

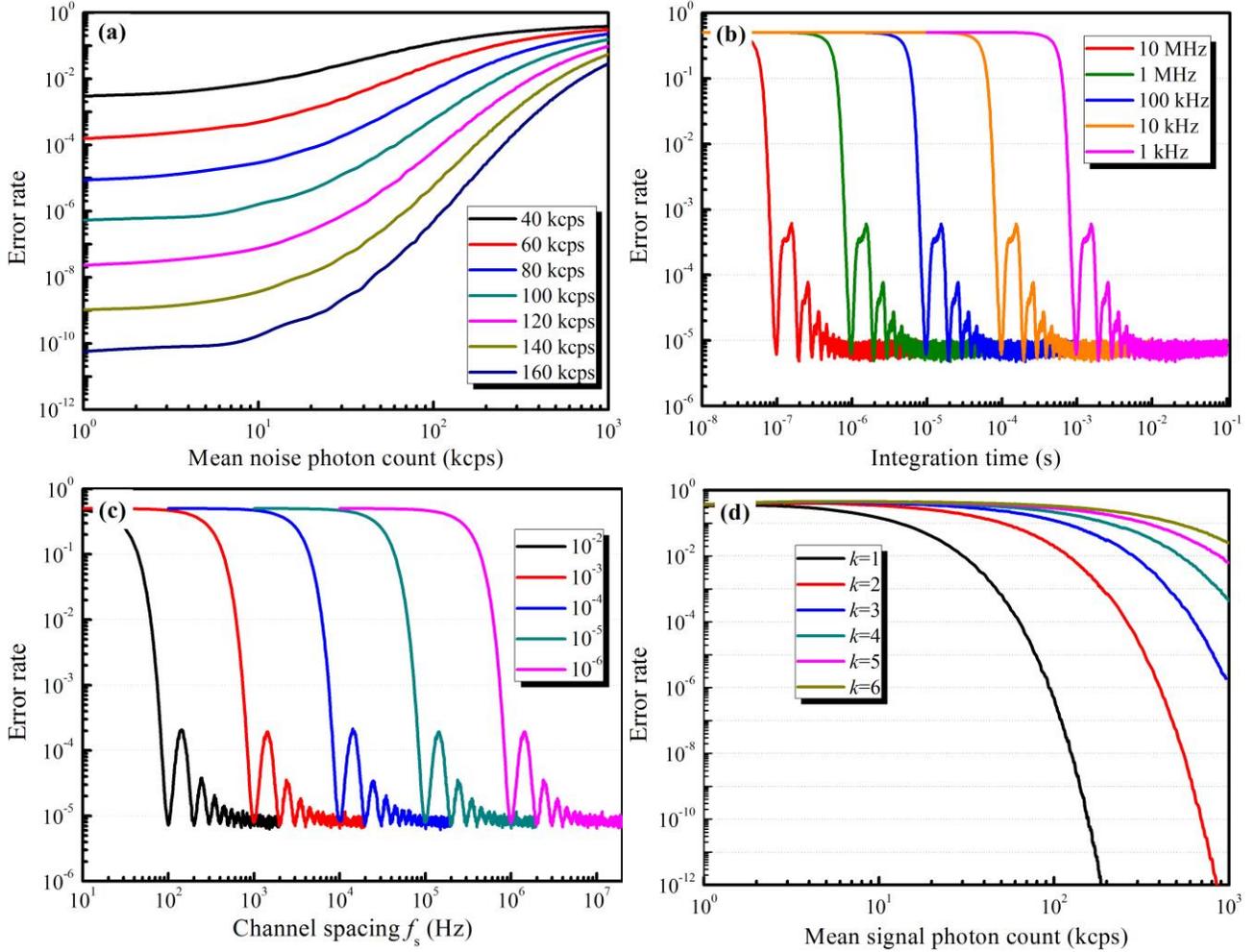

FIG. 5. (Simulation) (a) The error rate with different mean noise photon count. Each line represents different mean signal photon count. (b) Integration time dependence of the error rate of MCFC scheme. The error rate is oscillating with the increase of integration time, the period is $1/f_m$. The oscillation waveform is correlated with the initial phase $\varphi$ of the modulation signal. In this simulation, initial phase $\varphi=0$ and the mean signal photon count in each integration time is 80. (c) The error rate vs channel spacing. For different integration time the relationship between the error rate and the channel spacing is also different. In (a), (b) and (c) only one frequency component is considered. (d) Error rate changes with different frequency component, here $k$ ($k=1, 2, 3, 4, 5, 6$) is the number of frequency component.

the error rate is oscillating with integration time. In addition, the amplitude of the oscillation is fading with the increase of the integration time and tend to be a certain level. Further improvement of the error rate is limited by the quantum shot noise. In practical applications, the integration time should be as small as possible to obtain the highest transmission capacity, as described in Eq. (6). The higher the modulation frequency the shorter the integration time required. Therefore, when determining the integration time only the lowest modulation frequency used in the system need to be considered. In Fig. 5(b), as the dot line shows, if one set the threshold of the error rate at $10^{-5}$, the integration time would be set at 1μs, 10 μs, 100 μs, 1 ms and 10 ms when the minimum modulation frequency is 10 MHz, 1 MHz, 100 kHz, 10 kHz and 1 kHz, respectively.

The full width at half maximum of the characteristic spectrum line is mainly influenced by the integration time. Channel spacing should be set properly to avoid the disturbance of adjacent frequency channels. Fig. 5(c) shows the relationship between the error rate and the channel spacing for different integration time. The different color lines indicate different integration time. The error rate shows damped oscillation with the increase of the channel spacing. To optimize the transmission capacity, small channel spacing is preferred. Therefore, the first trough of each line is the best choose of the channel spacing, as the red line shows, where the integration time is 1 ms and the channel spacing prefer to set at 1 kHz.

Number of frequency component $k$ also has influence on the error rate. Figure 5(d) shows the changes of error rate for different frequency components. For a certain error rate, with the increase of frequency components more photon counts are needed. For example, when the error rate is set at $10^{-5}$, the detected mean signal photon count need to be 80 kcps or 720 kcps when there are one or three frequency components. The relationship between amplitude of the characteristic spectrum line and the number of frequency component is nonlinear, especially at lower mean signal photon count, see more detailed description in the Sec. S3 of the Supplemental Material [32]. This is because that with the increase of the frequency components the modulation signal is more sophisticated where need more photons to recover the modulation frequency information.

The requirement of error rate is different in various situations, usually, the error rate less than $10^{-5}$ is enough for information transmission [12, 35]. Figure 5 can be used as a reference for parameter selection for different requirement of error rate. When considering the error rate, the transmission capacity is

$$I = \frac{1}{T_I} \text{Log}_2 M_{\max} \left(1 - H(p_e)\right). \quad (8)$$

$p_e = M_{\max} \cdot e / 2(M_{\max} - 1)$ is the fale information ratio; and $H(p_e) = [-p_e \cdot Log_{M_{\max}} p_e - (1-p_e) \cdot Log_{M_{\max}}(1-p_e)] \cdot Log_2 M_{\max}$ is the shannon entropy. In the practical experiments, the time jitter of SPAD is approximately 400 ps, so the total available modulation bandwidth $B \approx 1$ GHz. Assume that $f_s = 1.0$ kHz, $T_I = 1$ ms, when frequency component $k$ is one or three, the transmission capacity $I$ is 19.2 kbps and 57.2 kbps, respectively. The error rate is determined by the photon number the receiver detected.

The MCFC coding scheme can be easily extended to the phase or polarization modulation from the intensity modulation, by using a phase modulator or a polarization modulator.

## V. CONCLUSION

In this paper, we proposed a new MCFC coding scheme, which is specially designed for laser communication when the received signal is at the single-photon level. In this scheme, the modulation frequency could be operated in a wide band without frequency aliasing due to the inherent randomness of photons arrival time of weak coherent light. A typical color image had been transmitted to verify its feasibility. The studies of statistical properties of the SPMS, including the dependence of mean noise photon count, integration time, channel spacing and the number of frequency component, helped us to optimize the system parameters. These works can be used as a reference for practical parameter choosing for different requirement of error rate. Finally, the transmission capacity is presented when considering the error rate.

The coding scheme can be used in the long-distance laser communications, such as the construction of interplanetary Internet. It is helpful in reducing the requirement of laser transmitting power and the size of receiving antenna. The MCFC scheme also shows great potential applications in single-photon laser ranging, quantum communication, and so on.

## ACKNOWLEDGMENTS

We thank the high performance simulation platform of Shanxi University to provide computing resource. The project is sponsored by the Natural Science Foundation of China (Nos. 61527824, 11374196, and 61675119) and PCSIRT (No. IRT 13076).

# Supplemental Material for "Enhanced transmission capacity for laser communication at the single-photon level using the multi-channel frequency coding scheme"

Jianyong Hu, Bo Yu, Mingyong Jing, Liantuan Xiao* and Suotang Jia

*State Key Laboratory of Quantum Optics and Quantum Optics Devices, Institute of Laser Spectroscopy, Shanxi University, Taiyuan 030006, China*
*Collaborative Innovation Center of Extreme Optics, Shanxi University, Taiyuan 030006, China;*
*\*Corresponding author E-mail address: xlt@sxu.edu.cn*

## S1. Single photons modulation

In the experiment, a Mach-Zehnder interferometer type optical intensity modulator was used for single photons modulation. As show in Fig. S1, $P_{in}$ and $P_{out}$ are the input and output of the modulator, respectively [1].

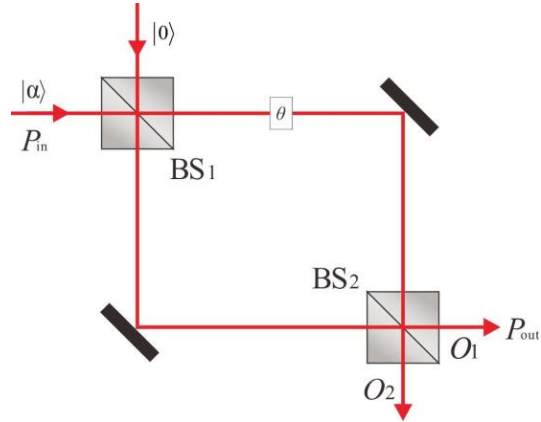

FIG. S1. Structure of Mach-Zehnder interferometer type optical intensity modulator.

For fock state $|n\rangle$, after the first beam splitter $BS_1$ it change to

$$|n\rangle \rightarrow \frac{1}{\sqrt{2}^n}(|1\rangle^T + i|1\rangle^R)^n . \tag{S1}$$

The photons transmit and reflect from the first beam splitter $BS_1$ are incident to the second beam splitter $BS_2$. Only $n_{out}$ ($n_{out} \leq n$) photons will pass through the modulator. After the $BS_2$ the input Fock state is changed to

$$|n\rangle \rightarrow \frac{1}{(\sqrt{2})^n}\sum_{n_{out}=0}^{n} i^{n_{out}} (\frac{n!}{n_{out}!(n-n_{out})!})^{\frac{1}{2}} (e^{i\theta}-1)^{n-n_{out}}(e^{i\theta}+1)^{n_{out}} |n_{out}\rangle_{O_1} |n-n_{out}\rangle_{O_2} . \tag{S2}$$

The output photon number of two output port is strongly dependent on the relative phase $\theta$.

$$P_{O_1} = \frac{n}{2}[1+\cos(\theta)],$$

$$P_{O_2} = \frac{n}{2}[1-\cos(\theta)]. \tag{S3}$$

If we change the relative phase $\theta$, the changed interval is within $0\sim\pi$, the two output ports of the second beam splitter $BS_2$ are changing with a cosine function.

The coherent state can be written as

$$|\alpha\rangle = e^{-|\alpha|^2/2}\sum_{N=0}^{\infty}\frac{\alpha^N}{(N!)^{1/2}}|n\rangle \tag{S4}$$

Take the coherent state as input state, the output state of the two ports is change to

$$P_{O_1} = \sum_{n=0}^{\infty} p_n \cdot \frac{|\alpha|^2}{2}\cdot[1+\cos(\theta)] ,$$

$$P_{O_2} = \sum_{n=0}^{\infty} p_n \cdot \frac{|\alpha|^2}{2} \cdot [1 - \cos(\theta)]. \tag{S5}$$

Here $p_n$ represents the probability of sending out $n$ photons, which obeys Poisson distribution. The output $P_{out}$ of the intensity modulator is equal to one of the output port, here we assume that $P_{out}=P_{O1}$. For a single-photon, $P_{out}$ can be seen as the probability to pass through the modulator. After the modulation, the distribution of the photon sequence is the superposition of the cosine distribution and the Poisson distribution [2, 3]. This can be seen from the measurement result of second-order-correlation function and Mandel Q parameter, as show in Fig. S2, the experimental data are acquired with the help of a time interval analyzer.

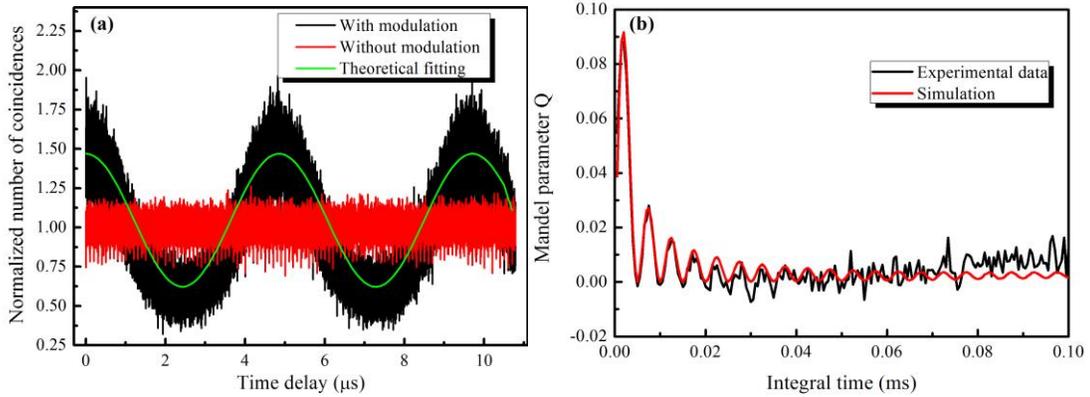

FIG S2. Experimental result of second-order-correlation function and Mandel parameter measurement. In figure (a), the black and red line indicate the second-order-correlation function of single-photon sequence with and without modulation. The green line is the fitting of the black line. In figure (b), the black and red line are experimental data and simulation result, respectively; the fluctuation of Mandel parameter is aggravated with the increase of the integral time, this is mainly caused by the limited sampling time.

## S2. Transmitting information using the MCFC scheme

The MCFC scheme can be used to transmit text, images, video and other file formats, we just need to build up the corresponding relationship between the files and the modulation frequencies. Furthermore, the total available modulation bandwidth could be divided into a plurality of sub-bands which are allocated for different users or different information types, and the relationship between information and modulation frequency in each of the sub-bands is respectively determined. For example, Table S1 shows a relationship between English characters and the modulation frequencies established by one user. The sub-band corresponding to the English characters is from 50 kHz to 75 kHz, the channel spacing is 1.0 kHz. In the same way, other users can build up the corresponding relationship between the English characters and the modulation frequencies in different sub-bands. The users encode or decode the information according to the corresponding relationship between the English characters and the modulation frequencies.

Table S1

| Character | A | B | C | D | E | F | G | H | I |
|---|---|---|---|---|---|---|---|---|---|
| Frequency (kHz) | 50 | 51 | 52 | 53 | 54 | 55 | 56 | 57 | 58 |
| Character | J | K | L | M | N | O | P | Q | R |
| Frequency (kHz) | 59 | 60 | 61 | 62 | 63 | 64 | 65 | 66 | 67 |
| Character | S | T | U | V | W | X | Y | Z |  |
| Frequency (kHz) | 68 | 69 | 70 | 71 | 72 | 73 | 74 | 75 |  |

## S3. Nonlinear effect of single photon modulation spectrum

With the increase of the mean signal photon count the amplitude and the variance of the single photons modulation spectrum increase in the non-linear form. As the inset of Fig. S3 (a), the amplitude of the noise floor increases nonlinearly, but the amplitude of characteristic spectrum line increases linearly. The variance of characteristic spectrum line and noise floor both increases nonlinearly. This is originated from quantum uncertainty principle which lead to the fluctuation of photon count.

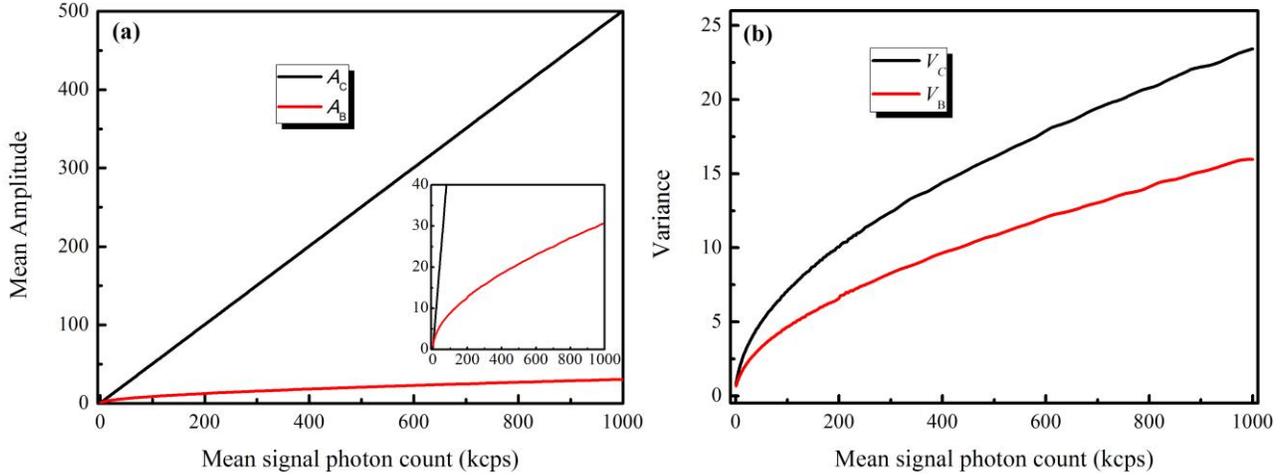

FIG S3. Nonlinear effect of single photon modulation spectrum. (a) $A_C$, and $A_B$ are mean amplitude of characteristic spectrum line and noise floor, respectively. (b) $V_C$ and $V_B$ are the variance of characteristic spectrum line and noise floor, respectively.

For a classical radio frequency signal, the amplitude of the characteristic spectrum line is inversely proportional to the number of frequency component, which means $A_1=kA_k$ ($k$=1, 2, 3, 4, 5, 6). However, at the single-photon level, the relationship between amplitude of the characteristic spectrum line and the number of frequency component is nonlinearly, especially at lower mean signal photon counts, because less digital sample could not recover the frequency spectroscopy completely [4, 5]. In the Fig. S4 (a), we simulated the mean amplitude of the characteristic spectrum line when the number of frequency component $k$=1, 2, 3, 4, 5, 6. The inset is the amplification of the area with the red dash line, it shows that the mean amplitude of characteristic spectrum line increases nonlinearly at low counting rate when the number of frequency component $k$ is bigger than one. With the increase of mean signal photon counts the modulation spectrum tend to be a classical spectrum, as show in Fig. S4 (b), with the increase of mean signal photon counts $A_1/A_2$, $A_1/A_3$, $A_1/A_4$, $A_1/A_5$ and $A_1/A_6$ tend to be 2, 3, 4, 5 and 6, respectively.

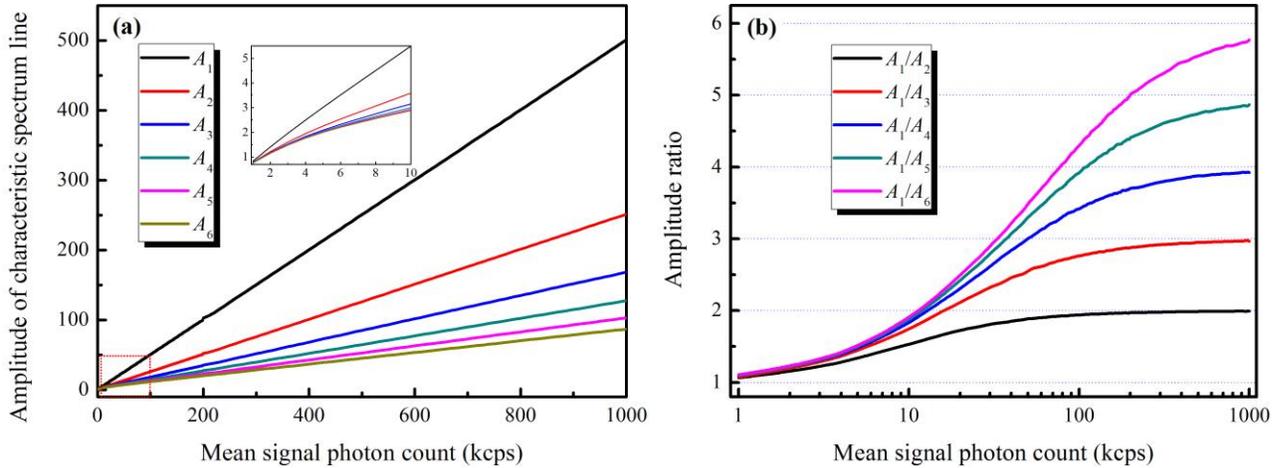

FIG S4. (a) Amplitude of characteristic spectrum line with different number of frequency components. $A_k$ ($k$=1, 2, 3, 4, 5, 6) is the amplitude of characteristic spectrum line, $k$ is the number of frequency components. (b) Amplitude ratio of characteristic spectrum line with different frequency component.